# Demand and Price Fluctuations Effect on Risk and Profit of Single and Clustered Microgrids during COVID-19 Pandemic

Tohid Khalili[1, *], Ali Bidram[1], Janie M. Chermak[2]

[1] Department of Electrical and Computer Engineering, University of New Mexico, Albuquerque, USA

[2] Department of Economics, University of New Mexico, Albuquerque, USA

*Corresponding Author

Emails: khalili@unm.edu, bidram@unm.edu, jchermak@unm.edu

**Abstract**— COVID-19's widespread distribution is wreaking havoc on people's lives all over the world. This pandemic has also had a significant impact on energy consumption. Its influence can be seen in the power system's operation and the market as well. The power consumers' habits and demand curves have been changed at a breakneck pace. In this work, a one-year mixed-integer programming (MIP) problem has been developed to compare the power consumption between 2019 and 2020 in the United States as an example regarding the COVID-19 pandemic effect in order to better prepare for possible similar future events. 100% renewable single microgrids (SMGs) are studied using wind turbines and photovoltaics. Batteries are also employed since it is inevitable when the system uses renewables. Additionally, it is possible for the SMGs to trade power with the main grid as needed. The effect of the SMGs' clustering to form the multi-microgrids (MMGs) is also considered. In order to investigate the risk of the system during the COVID-19 and formation of MMG, downside risk constraints are applied to the proposed model. Furthermore, a stylized short-run consumers demand model is proposed, using elasticity and assessed responses regarding the average household consumption for households during on-peak and off-peak periods. The simulation results show that COVID-19 generally reduces the demand, increases the profit of the system, and decreases the economic risk of the power system's operation. Moreover, SMGs clustering to organize MMG dramatically enhances the profit of the system as well as improves the risk level of the system.

**Acronyms**

| Acronym | Description |
|---|---|
| EIA | The U.S. Energy Information Administration |
| ERCOT | Electric Reliability Council of Texas |
| GAMS | General Algebraic Modelling System |
| GC | Grid-connected |
| MG | Microgrid |
| MIP | Mixed-integer programming |
| MMG | Multi-microgrid |
| NREL | National Renewable Energy Laboratory |
| PV | Photovoltaic |
| RES | Renewable energy source |
| SMG | Single microgrid |
| TOU | Time of use |
| U.S. | United States |
| WHO | World Health Organization |
| WT | Wind turbine |

**Nomenclature**

| | |
|---|---|
| **Indices** | |
| $h_m$ | number of hours in each month |
| $z$ and $y$ | alias indices for the number of the SMGs ( $z,y=1,2,3$ ) |
| m | month |
| $t$ | time |
| **Parameters and Variables** | |
| $\frac{\partial PL}{\partial C}$ | slope of the demand function and $C$ and $PL$ |
| $\varepsilon$ | demand elasticity |
| $C_{t,m}^{buy}$ | price of the power bought from the SMGs or the main grid |
| $fz$ | considered number of SMGs |
| $CVD_m$ | average demand change between 2020 and 2019 |
| $P_{zy}^{linemax}$ | maximum capacity of the lines |
| $X_{zy,t,m}^{grid}$ | a binary variable toggling between buying and selling |
| $S_z$ | base value of the power |
| $W_{z,m}$ | a binary variable |
| $t\,arg\,et_{z,m}$ | considered target |
| $\overline{EDR}_{z,m}$ | expected downside risk |
| $OF_{1,m}$ , $OF_{2,m}$ , $OF_{3,m}$ | profit of the SMG1, SMG2, SMG3 |
| $C_{t,m}^{sell}$ / $C_{t,m}^{sell-grid}$ | price of the power sold to the SMGs / the main grid |
| $P_{22,t,m}^{sell}$ / $P_{22,t,m}^{buy}$ | SMG2's sold/purchased power to/from the main grid |
| $P_{21,t,m}^{sell}$ / $P_{23,t,m}^{sell}$ | sold power to the SMG1 / SMG3 by the SMG2 |
| $P_{21,t,m}^{buy}$ / $P_{23,t,m}^{buy}$ | purchased power from the SMG1 / SMG3 by the SMG2 |
| $P_{33,t,m}^{sell}$ / $P_{33,t,m}^{buy}$ | SMG3's sold/purchased power to/from the main grid |
| $P_{31,t,m}^{sell}$ / $P_{32,t,m}^{sell}$ | sold power to the SMG1 / SMG2 by the SMG3 |
| $P_{31,t,m}^{buy}$ / $P_{32,t,m}^{buy}$ | purchased power from the SMG1 / SMG2 by the SMG3 |
| $PWT_{z,t,m}^{max}$ / $PPV_{z,t,m}^{max}$ | maximum generation of the WTs/PVs |
| $P_z^{batmin}$ / $P_z^{bat\,max}$ | minimum and maximum power of the batteries |
| $SOC_z^{min}$ / $SOC_z^{max}$ | Minimum / maximum state of charge of the batteries |
| $SOC_{z,t,m}$ | state of charge of the batteries |
| $PL_{1,t,m}$, $PPV_{1,t,m}$, $PWT_{1,t,m}$, and $P_{1,t,m}^{bat}$ | load, PV generation, WT generation, and battery charge/discharge of the SMG1 |
| $PL_{2,t,m}$, $PPV_{2,t,m}$, $PWT_{2,t,m}$, and $P_{2,t,m}^{bat}$ | load, PV generation, WT generation, and battery charge/discharge of the SMG2 |
| $PL_{3,t,m}$, $PPV_{3,t,m}$, $PWT_{3,t,m}$, and $P_{3,t,m}^{bat}$ | load, PV generation, WT generation, and battery charge/discharge of the SMG3 |
| $P_{11,t,m}^{sell}$ / $P_{11,t,m}^{buy}$ | SMG1's sold/purchased power to/from the main grid |
| $P_{12,t,m}^{sell}$ / $P_{13,t,m}^{sell}$ | sold power to the SMG2 / SMG3 by the SMG1 |
| $P_{12,t,m}^{buy}$ / $P_{13,t,m}^{buy}$ | purchased power from the SMG2 / SMG3 by the SMG1 |
| $A$ | Intercept point |
| $B$ | other relevant factors |
| $C$ | price |
| $fm$ | number of months |
| $M$ | a large and positive number |
| $OF^{total}$ | total profit over $fm$ months |
| $PL$ | quantity demanded at a point in time |
| $risk^{total}$ | total risk of the system |
| $risk_{z,m}$ | risk of the system |

## 1. Introduction

In the last days of 2019, the COVID-19 pandemic began. According to the World Health Organization (WHO) statistics, approximately 240 million individuals were infected with COVID-19, and around 5 million reported fatalities were determined owing to the COVID-19 through October 13, 2021 [1]. COVID-19 changed people's lives in numerous ways. One of the effects of the COVID-19 pandemic was a dramatic shift in global energy output and demand [2]. Due to the COVID-19 outbreak, global power demand has significantly decreased [3]. For example, in the United Kingdom, power demand normally drops by 10-20% on weekends compared to weekdays; but, since COVID-19, the same drop in demand has been observed throughout the week. Another case is the 2.88% decrease in power consumption in the United States (U.S.) in 2020 compared to 2019. These unusual events impacted electricity production and trade.

On the other hand, power grids are continuously exposed to a variety of unforeseen and extremely unknown risks that might jeopardize the long-term sustainability, reliability, and supply of the energy required to light up present civilization [4]. As a result, the consequences of the COVID-19 pandemic on the power systems have been studied in several recent publications [5-9]. Monitoring the load fluctuations is critical for balancing the generation and demand in order to ensure the power systems' reliability [10]. Reference [11] studied an early view about the effect of the COVID-19 pandemic on the United States generation and load. In addition, the impact of the pandemic on the energy demanded by consumers is analyzed in the case of Northern Italy [12]. In [13], power demand and supply changes are analyzed under the lockdown circumstances during the COVID-19 pandemic considering the stability of the power system. Due to the fact that consumers' habits and load patterns changed at a phenomenal rate, [14] and [15] investigated the impacts of the COVID-19 pandemic on the power systems of Terni (a city in Italy) and Bangladesh, respectively. A complete review of the United States market and power systems activities is conducted in [16] considering the COVID-19 crisis by developing a data-driven analysis to uncover the effects of COVID-19 on the power demand and supply. Reference [17] performed a long-term prediction-based assessment on the power demand gap affected by COVID-19 at a case in China. The impact of the pandemic on the local distribution transformers and residential demand is evaluated in [18].

Multi-microgrids (MMGs) have been introduced as a promising solution to increase the resilience of power grids. In [19], an electricity market approach for the energy management of MMGs is proposed. Reference [20] presents a peer-to-peer selling and buying method for a cluster of microgrids (MGs) using a multi-objective optimization approach. Optimal management of renewable energy sources (RESs) with batteries in the distribution systems is offered in [21]. On the other hand, some references have addressed the utilization of single microgrids (SMGs). In [22], the economic and optimal performance of SMGs is analyzed in order to enhance the resilience of the system considering the unpredictable outages. In [23], stochastic management of a renewable SMG using the demand response program is addressed. An optimization method is presented for scheduling an SMG using renewable sources in a multi-objective manner to satisfy the consumers [24].

In MGs with RESs, the price of the power purchased from utility is usually lower than the RESs' power price. Moreover, the intermittent nature of RESs imposes more risk on profit of the MG operator (i.e., MG operator may gain less profit from the desired profit values). Reference [25] investigated the effect of the risk on the profit of SMGs with the demand response program by presenting an optimization approach. Additionally, the risk of a resilient renewable-based system is explored by sharing the risk during the COVID-19 [26]. In order to find a proper demand model, there are numerous elasticity estimates in the extant literature. For instance, [27] and [28] provide a synopsis of estimated short-run elasticities.

This research assesses the impact of COVID-19 on the load and risk of SMGs and MMG over a year in order to improve the profit of the system. The demand of the considered grid-connected (GC) system is supplied by local wind turbines (WTs) and photovoltaic (PV) systems and trading with the main grid. The proposed mixed-integer programming (MIP) problem is solved by employing a demand model based on local surveys. The main contributions of this paper are explained as follows:

- Annual optimal scheduling is performed to find the impact of clustering GC SMGs to form an MMG under high penetration of RES including WTs and PVs and batteries. The system's local generators are 100% renewable, and they are supported by the main grid.
- A stylized short-run demand and price relation model is utilized for electricity, incorporating estimated response considering the elasticity.
- Economical risk of the SMGs and MMG during the COVID-19 pandemic is investigated. Thus, the downside risk constraints are applied. Also, the system's profit is optimized and compared in each architecture over a year. The obtained simulation results show the decrement of the risk at SMGs and MMGs. Moreover, the results show that by clustering the SMGs and forming the MMG, the profit of the system increases the system's risk decreases.
- The effect of the COVID-19 on all studied architectures is examined during 2020 and the obtained results are compared with the 2019 data in which they were not affected by the COVID-19 pandemic. The results of studies show that since the COVID-19 reduces the demand for the power system, the profit of both SMGs and MMGs when considering the pandemic's effect increases.

The remainder of this paper is organized as follows: Section 2 presents the proposed studied model including the architecture, components, and parameters. Section 3 discusses the problem formulation by proposing the objective functions and constraints. The obtained simulation results and discussion are provided in Section 4. Finally, the conclusions are listed in Section 5.

## 2. System Model and Data

### *2.1. System Architecture*

The system consists of three SMGs. This model can be used in residential cases, distribution systems, microgrids as well as local and remote communities regardless of the size of the system and the number of prosumers. With a greater number of prosumers, more profit more profit can be achieved by the operator and customers. The proposed model is fully scalable and does not depend on the number of SMGs in the clustered microgrid structure. When the demand is low, the excess of generated power of the SMGs can be sold to the main grid and if the generation is lower than the demanded load, power can be purchased from the main grid. The system has spilled energy due to the line capacity and the maximum selling amount of power to the main grid. Moreover, battery capacity can be used to compensate the mismatches between generation and load. However, when the battery is fully charged and there are excess generation and restriction on selling power, the system will have spilled energy. On the other hand, when the load is too high, the operator may need to use the load shedding methods to reduce the load. Based on the interconnection of three SMGs, two architecture are considered which are elaborated as follows:

### *2.1.1. First Architecture (GC MMG)*

In the first architecture, SMGs are operated individually; they are connected to create an MMG. Furthermore, it is possible to trade (buy/sell) power with the main grid and among SMGs as demonstrated in Figure 1.

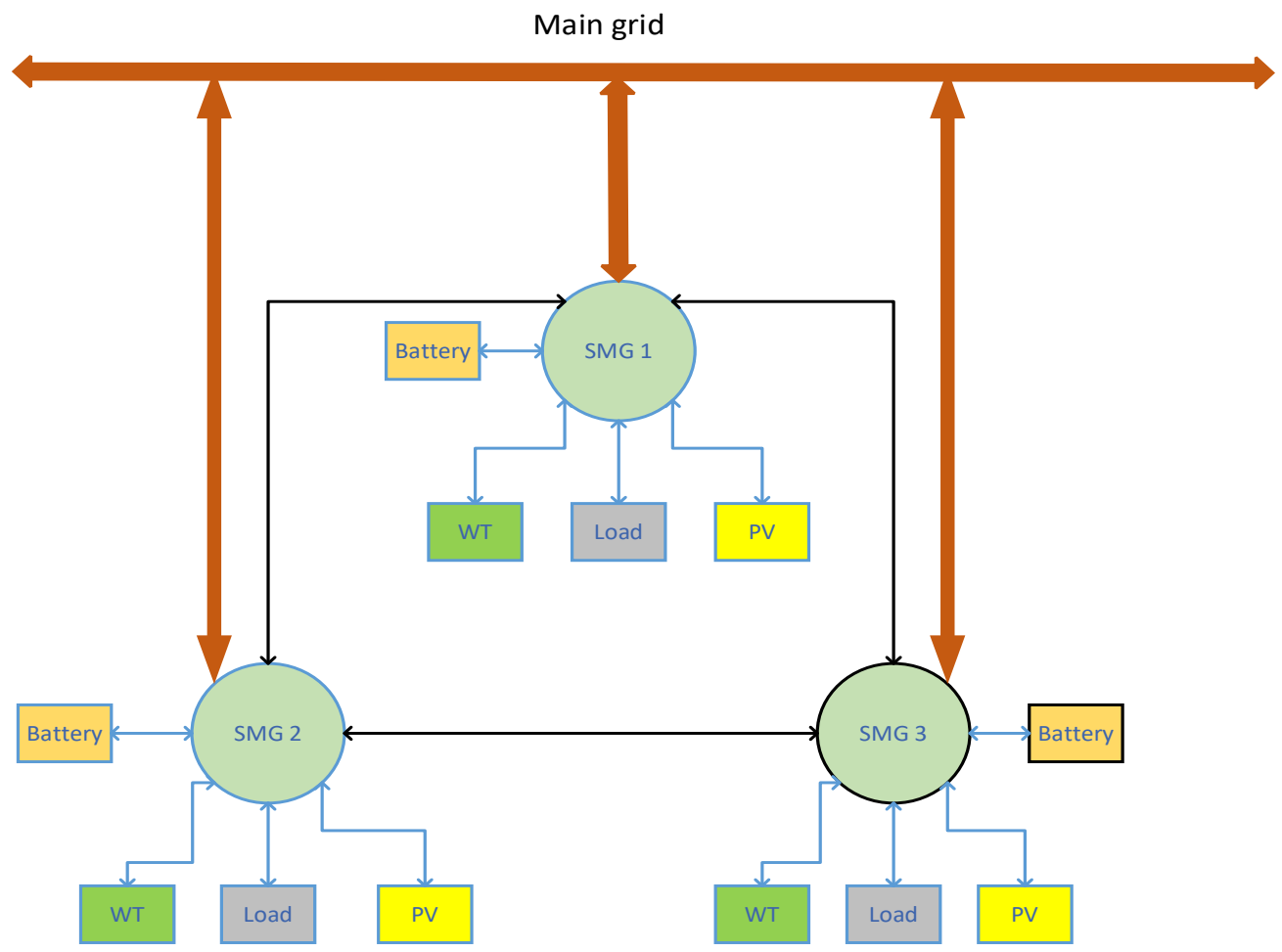

Figure 1. Schematic of the first architecture.

### *2.1.2. Second Architecture(GC SMGs)*

In the second architecture, the three studied SMGs are operating separately and all of them are connected to the main grid. Each SMG can individually trade power with the main grid as shown in Figure 2.

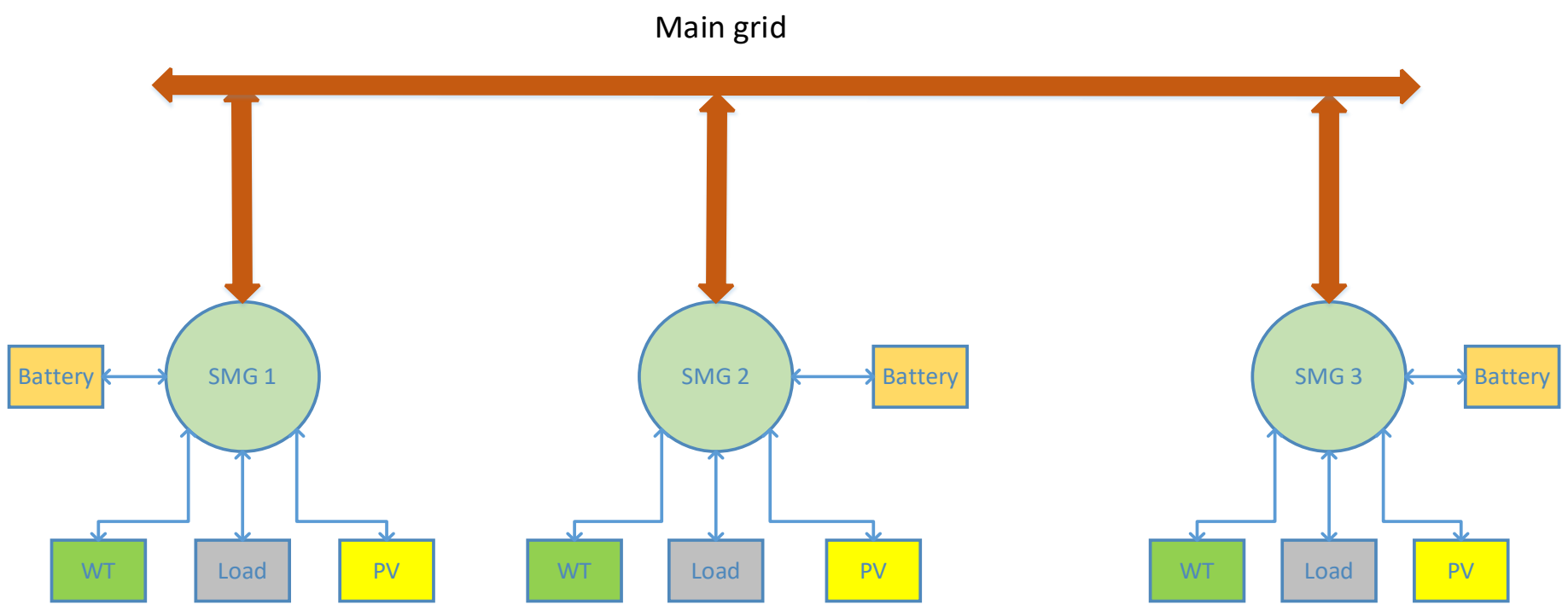

Figure 2. Schematic of the second architecture.

### *2.2. Input Data*

This paper utilizes the actual hourly load profile of the Electric Reliability Council of Texas (ERCOT) that is scaled down for the proposed system in Section 2.1 [29]. The size of each power source is determined based on the demand of the system in order to satisfy the demand. In all sources, we have used real data. The effective 8-hour sun pattern can be a challenge for supplying the load. Similarly, WT's generation depends on wind speed. To address these challenges, the SMGs are equipped with batteries to compensate for the mismatches between generation and load. In the studied system, the main grid pays 10% incentives when the SMGs sell power to the main grid in order to encourage SMGs owners as they are 100% RES-based. SMG1, SMG2, and SMG3 have high, average, and low demand profiles, respectively. SMG1's load is as twice as SMG2's load, and SMG2's load is as twice as SMG3's load. The aggregated hourly load profile of the whole system over a year is shown in Figure 3.

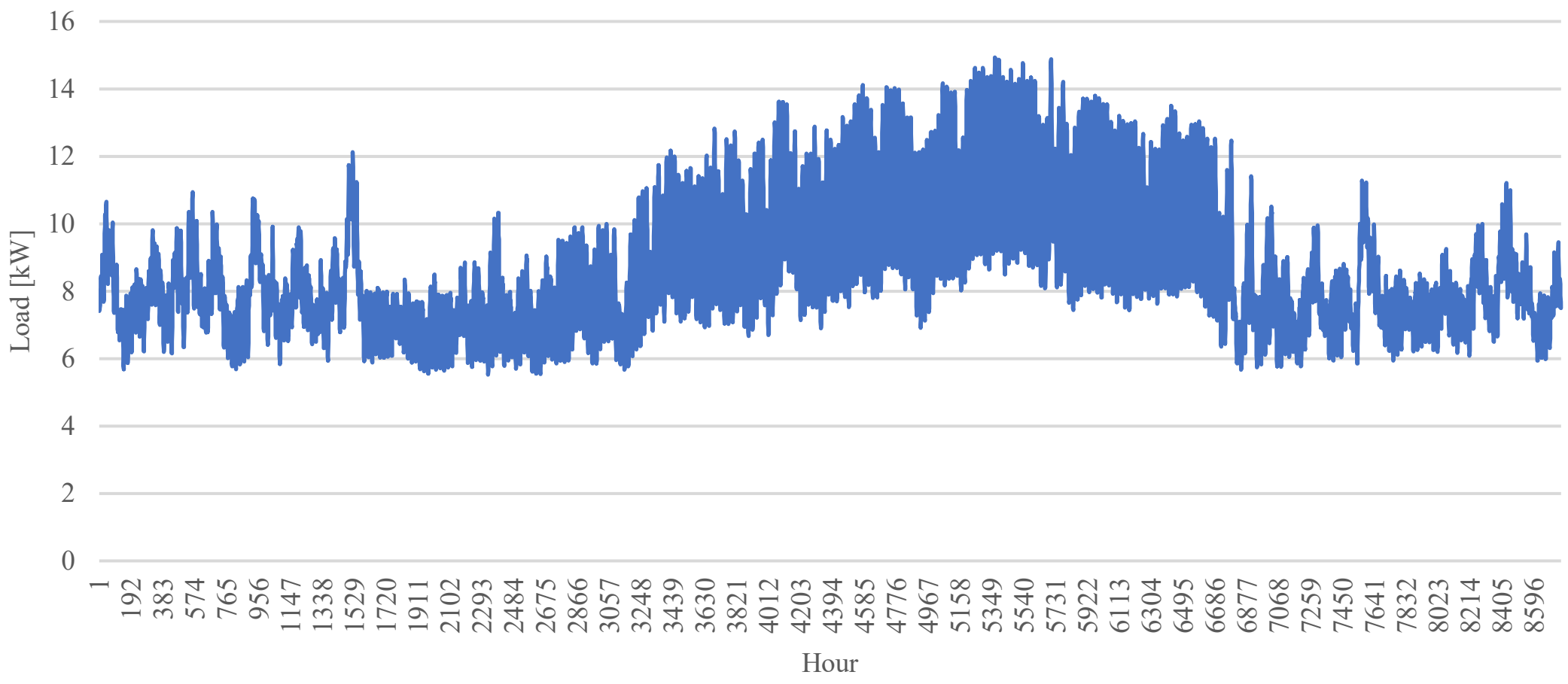

Figure 3. Total hourly load profile.

*Photovoltaic System Data:* The 8760 hourly PV profile is derived from the National Renewable Energy Laboratory (NREL) Solar power database [30]. This data consists of 1 year of 5-minute solar power and is dependent on different parameters, i.e., irradiance, location, and capacity of the system. The specific data for the state of New Mexico is used in this study. For our considered system, we have supposed that about 40 % ( ~ 32500 kWh) of the generation is generated by PVs. Then, the data of the NREL is scaled down to fulfill this assumption considering the demand of the considered system. Figure 4 illustrates the aggregated hourly profile of all of the PVs in the system. For each SMG, the PV profile in Figure 4 is proportionally scaled down based on the maximum demand of that SMG. The rated powers of the PVs are 12, 6, 3 kW

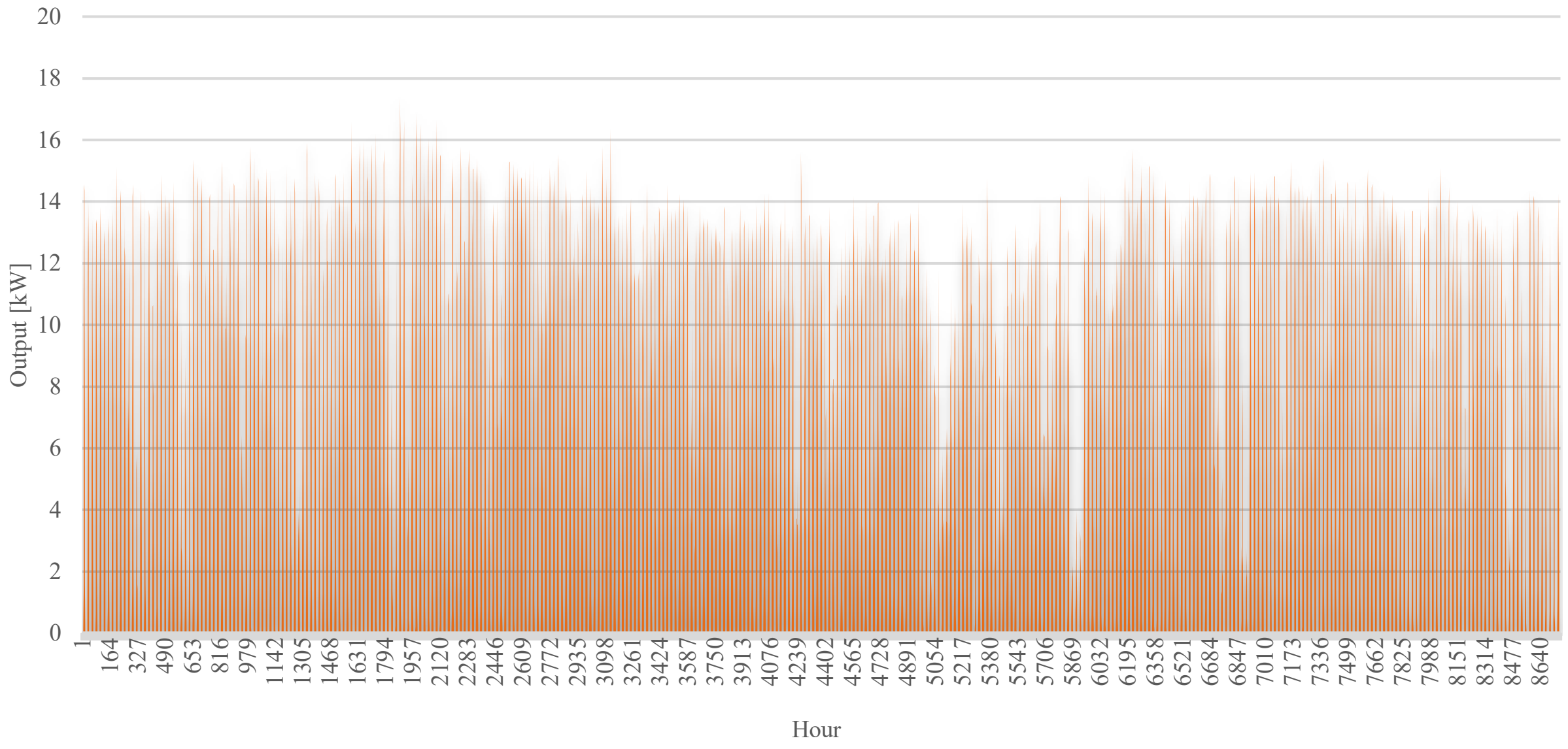

Figure 4. Aggregated PV profile.

*Wind Turbine Generator Data:* The 8760 hourly WT profile is derived from ERCOT [31]. The WT profile is scaled down to fit the requirements of the studied system. The intended WTs are Type 3 WT generators. This data is based on different parameters, i.e., daily and all-time peak wind generation and penetration. WTs generate about 60% ( ~ 48750 kWh) of the total generation of the system. Therefore, the data of the ERCOT has been scaled down to fulfill this assumption considering the considered system's demand. Figure 5 illustrates the aggregated hourly profile of all of the WTs in the system. For each SMG, the WT profile in Figure 5 is proportionally scaled down based on the maximum demand of that SMG. The rated powers of the WTs are 10, 5, and 3 kW.

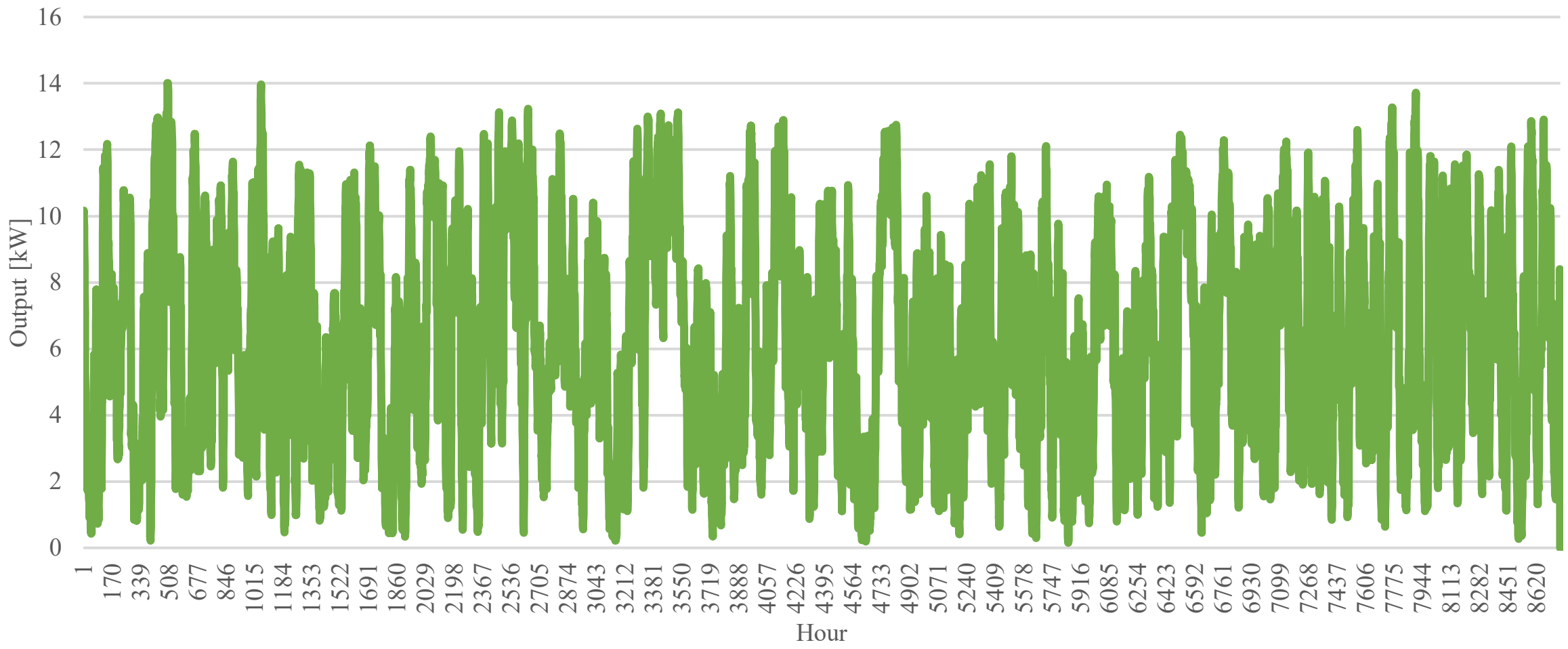

Figure 5. Aggregated WT profile.

*COVID-19 Data:* The U.S. Energy Information Administration (EIA) data about the load changes during 2020 compared to 2019 is used to model the effect of the COVID-19 pandemic on the studied model's variables and objective functions [32]. In 2020, The net generation of the U.S. is decreased mainly due to the mitigation efforts for the COVID-19 compared to 2019 [32]. Indeed, using the data obtained from the COVID-19 effect can help operators to be ready for future events regarding the load level and load profile during the massive events.

## 3. Problem Formulation

In this section, the mathematical formulation of the proposed model including the demand model, objective functions, and constraints are provided.

### *3.1. Stylized Short-Run Demand Model for Electricity*

Demand for electricity is the relationship between the quantity *PL*, *C*, and *B* and can be formulated as

$$PL = f(C, B). \tag{1}$$

It is, however, a derived demand, as consumers are not buying electricity to consume it directly, but rather electricity is consumed because of the goods and services it provides as an energy source, e.g., heating and cooling or lighting. Thus, the consumption of electricity depends on the demand for those goods and services and the other factors can be of significant importance. For residential consumption, that demand varies across the day, resulting in time of use (TOU) consumption patterns. Modeling electricity demand can be complex, as the variations across consumers and across time can result in heterogeneous consumption patterns, depending on consumer characteristics, and time-dependent consumption.

Elasticity is the responsiveness of demand to a change in another factor. Demand elasticity is the responsiveness of demand given a change in price and is estimated by

$$\varepsilon = \frac{\partial PL}{\partial C}\frac{C}{PL}. \tag{2}$$

The interpretation of elasticity depends on the absolute value of the measure, where

$$\begin{cases} |\varepsilon| > 1 \text{ elastic} \rightarrow \%\boldsymbol{\Delta} PL > \%\boldsymbol{\Delta} C \\ |\varepsilon| = 0 \text{ unit elastic } \rightarrow \%\boldsymbol{\Delta} PL = \%\boldsymbol{\Delta} C \\ |\varepsilon| < 1 \text{ inelastic } \rightarrow \%\boldsymbol{\Delta} PL < \%\boldsymbol{\Delta} C \end{cases}. \tag{3}$$

This paper develops a stylized short-run demand model for electricity, incorporating estimated response, or elasticity measures from the literature and average household consumption for households (captured by surveys) during on-peak and off-peak periods. The obtained data is modified to be appropriate to the scale of the proposed SMGs. The other relevant factors, *B*, include on-peak, off-peak, and tiered usage levels.

Considering the studies performed in the U.S., the average elasticity values in individual studies range from -0.44 to -0.08 [33], depending on whether the estimate is a short or long run. That is, the percentage change in the quantity demanded is less than the percentage change in price, thus, consumers are relatively unresponsive to price change. We base our responses on [33], using a range from -0.16 to -0.08 for elasticity.

In order to develop a suite of demand functions, we assume a linear function,

$$PL = A - \frac{\partial PL}{\partial C}C, \tag{4}$$

where *A* is the intercept and $\frac{\partial PL}{\partial C}$ is the slope, which can be estimated utilizing an elasticity, *C*, and *PL*. That is

$$\frac{\partial PL}{\partial C} = e\frac{PL}{C}. \tag{5}$$

We utilize the estimated elasticity measures, with EIA average New Mexico residential monthly demand and price (over the 2017-2019 time period) to estimate the slope and corresponding intercept, *A*, (consistent with that slope, monthly demanded quantity, and price) to develop TOU demand functions for an "average" consumer and for customers plus or minus two standard deviations from the average (for the 2017-2019 data), resulting in a total of 10 potential demand functions per month.

*3.2. Objective Functions*

The main objective function of this paper is to maximize the profit of the system under different architectures and conditions over a year. Moreover, the optimal output of sources and traded power are calculated. The optimization is performed to obtain the transferred power among the SMGs and main grid depending on the architecture under study (i.e., two architectures introduced in Section 2.1). Optimal scheduling is implemented to find the power generated by the WTs and PVs, and the charge and discharge schedule of the batteries.

#### *3.2.1. First Architecture*

For the proposed model in Figure 1, the objective functions are formulated to maximize the profit of SMG1, SMG2, SMG3, and MMG, respectively. The objective functions consist of the revenue from selling power and the cost of buying power in each SMG. The MMG's profit is found by aggregating the individual SMG profits. The objective functions for month *m* are formulated as

$$OF_{1,m} = \sum_{t=1}^{h_m}\left(\left(P_{11,t,m}^{sell} * C_{t,m}^{sell-grid}\right)+\left(P_{12,t,m}^{sell}+P_{13,t,m}^{sell}\right)*C_{t,m}^{sell}\right)-\sum_{t=1}^{h_m}\left(\left(P_{11,t,m}^{buy}+P_{12,t,m}^{buy}+P_{13,t,m}^{buy}\right)*C_{t,m}^{buy}\right), \tag{6}$$

$$OF_{2,m} = \sum_{t=1}^{h_m}\left(\left(P_{22,t,m}^{sell} * C_{t,m}^{sell-grid}\right)+\left(P_{21,t,m}^{sell}+P_{23,t,m}^{sell}\right)*C_{t,m}^{sell}\right)-\sum_{t=1}^{h_m}\left(\left(P_{22,t,m}^{buy}+P_{21,t,m}^{buy}+P_{23,t,m}^{buy}\right)*C_{t,m}^{buy}\right), \tag{7}$$

$$OF_{3,m} = \sum_{t=1}^{h_m}\left(\left(P_{33,t,m}^{sell} * C_{t,m}^{sell-grid}\right)+\left(P_{31,t,m}^{sell}+P_{32,t,m}^{sell}\right)*C_{t,m}^{sell}\right)-\sum_{t=1}^{h_m}\left(\left(P_{33,t,m}^{buy}+P_{31,t,m}^{buy}+P_{32,t,m}^{buy}\right)*C_{t,m}^{buy}\right), \tag{8}$$

$$OF^{total} = \sum_{m=1}^{fm}\left(OF_{1,m}+OF_{2,m}+OF_{3,m}\right). \tag{9}$$

$C_{t,m}^{sell-grid}$ is 10% more than the price of the power in the internal trade of the system due to incentives paid by the main grid to the system.

#### *3.2.2. Second Architecture*

The second architecture is based on the proposed model in Figure *2*. In this architecture, the variables that are used for the internal trade of power are not considered in the objective functions, i.e.,

$$if\ z \neq y\ \ P_{zy,t,m}^{sell}\ \&\ P_{zy,t,m}^{buy}=0 \quad . \tag{10}$$

### *3.3. Constraints*

The studied system and four scenarios have several constraints with and without considering the COVID-19 effect that are explained as follows:

#### *3.3.1. First Architecture Considering the COVID-19*

In the first scenario, the power balance equations for the first architecture regarding the COVID-19 effect on the demand of the SMGs are given as

$$CVD_m * PL_{1,t,m} = PPV_{1,t,m} + PWT_{1,t,m} + P^{bat}_{1,t,m} + P^{buy}_{11,t,m} - P^{sell}_{11,t,m} + P^{buy}_{12,t,m} - P^{sell}_{12,t,m} + P^{buy}_{13,t,m} - P^{sell}_{13,t,m}, \quad (11)$$

$$CVD_m * PL_{2,t,m} = PPV_{2,t,m} + PWT_{2,t,m} + P^{bat}_{2,t,m} + P^{buy}_{22,t,m} - P^{sell}_{22,t,m} + P^{buy}_{21,t,m} - P^{sell}_{21,t,m} + P^{buy}_{23,t,m} - P^{sell}_{23,t,m}, \quad (12)$$

$$CVD_m * PL_{3,t,m} = PPV_{3,t,m} + PWT_{3,t,m} + P^{bat}_{3,t,m} + P^{buy}_{33,t,m} - P^{sell}_{33,t,m} + P^{buy}_{31,t,m} - P^{sell}_{31,t,m} + P^{buy}_{32,t,m} - P^{sell}_{32,t,m}, \quad (13)$$

where $CVD_m$ denotes the percentage of the change of 2020 average load in month *m* compared to the average load of a similar month in 2019 [32].

Also, each line has its power transfer limit in the system; the limits of the interconnecting lines are defined as

$$0 \le P^{buy}_{zy,t,m} \le P^{linemax}_{zy} * X^{grid}_{zy,t,m} \quad , \quad (14)$$

$$0 \le P^{sell}_{zy,t,m} \le P^{linemax}_{zy} * \left(1 - X^{grid}_{zy,t,m}\right) \quad . \quad (15)$$

$X^{grid}_{zy,t,m}$ is set equal to one when the power is purchased, and $X^{grid}_{zy,t,m}$ is set equal to zero when the power is sold. The power bought from SMG1 by SMG2 is the same as the power sold to SMG2 by SMG1. So, the following constraint is also considered for all SMGs' power trades

$$if\ zy{=}yz\ \&\ z \ne y \quad P^{sell}_{zy,t,m} = P^{buy}_{yz,t,m}. \quad (16)$$

The power generation limitation of the WTs and PVs are given as

$$0 \le PWT_{z,t,m} \le PWT^{max}_{z,t,m}, \quad (17)$$

$$0 \le PPV_{z,t,m} \le PPV^{max}_{z,t,m}. \quad (18)$$

The employed batteries' constraints are defined as

$$P^{batmin}_{z} \le P^{bat}_{z,t,m} \le P^{bat\,max}_{z}, \quad (19)$$

$$SOC^{min}_{z} \le SOC_{z,t,m} \le SOC^{max}_{z}, \quad (20)$$

$$SOC_{z,t,m} = SOC_{z,t-1,m} - P^{bat}_{z,t,m} / S_z, \quad (21)$$

where (21) calculates the $SOC_{z,t,m}$ with $S_z$ defined as the base value of the power.

In order to calculate the risk of the profit, the downside risk method [25] is applied to the model. Downside risk constraints are defined as

$$0 \le risk_{z,m} \le M * W_{z,m}, \quad (22)$$

$$0 \le risk_{z,m} - \left(target_{z,m} - OF_{z,m}\right) \le M * \left(1 - W_{z,m}\right), \quad (23)$$

$$risk_{z,m} \le \overline{EDR}_{z,m}, \quad (24)$$

$$risk^{total} = \sum_{z=1}^{fz}\sum_{m=1}^{fm} risk_{z,m} \ . \tag{25}$$

Equation (22) shows the allowable range of risk. $risk_{z,m}$ is obtained regarding the obtained profit. Equation (23) calculates the downside risk regarding $target_{z,m}$, and the obtained profit of the system. Equation (24) shows the acceptable range of the downside risk regarding $\overline{EDR}_{z,m}$. Equation (25) calculates $risk^{total}$, over a year.

### *3.3.2. First Architecture without Considering the COVID-19*

In the second scenario, $CVD_m$ 's value is set equal to 1 and all other constraints are the same as *Section 3.3.1.*

### *3.3.3. Second Architecture Considering the COVID-19*

In the third scenario, all the constraints of the *Section 3.3.1* are considered as well as (10).

### *3.3.4. Second Architecture without Considering the COVID-19*

In the fourth scenario, $CVD_m$ 's value is set equal to 1 and all other constraints are the same as *Section 3.3.1.* Also, (10) is employed as well.

## 4. Results and Discussion

The optimization problem in Section 3 is an MIP problem that is modeled in the General Algebraic Modelling System (GAMS). This MIP model is solved by CPLEX solver. CPLEX solver is a powerful solver in GAMS to solve large-scale linear, quadratically constrained, and mixed-integer programming optimization problems. The type of problems that can be solved by CPLEX includes linear programming (LP), mixed integer programming (MIP), mixed integer programs with quadratic terms in the constraints (MIQCP), Quadratically constrained quadratic program (QCQP), Relaxed mixed integer programming (RMIP), Relaxed mixed-integer quad. constrain program (RMIQCP) [34]. In this work, four different scenarios are investigated which are as follows:

1) MMG (first architecture) with COVID-19 effect
2) MMG (first architecture) without COVID-19 effect
3) SMG (second architecture) with COVID-19 effect
4) SMG (second architecture) without COVID-19 effect

Table 1 summarizes the number of hours in each month of the year, the average monthly changes of the load profile of the U.S. in 2020 compared to 2019, and the corresponding $CVD_m$ for that month. It should be noted that from Table 1 data, the average reduction of the demanded load in 2020 compared to 2019 is 2.88%. As seen in Table 1, during the majority of the months, the load in 2020 has dropped compared to 2019. However, the load profile has slightly increased in February, June, July, and December due to the special occasions that happened in these months. The constraint parameters that are used in this research are provided in Table *2*.

Table 1. Total number of hours in each month and COVID-19 effect on the load.

| **Month** | $h_m$ | Change in Load from 2019 to 2020 [%] | $CVD_m$ |
|---|---|---|---|
| January | 744 | -5.20 | 0.948 |
| February | 672 | 1.00 | 1.01 |
| March | 744 | -5.60 | 0.944 |
| April | 720 | -6.70 | 0.933 |
| May | 744 | -7.60 | 0.924 |
| June | 720 | 0.40 | 1.004 |
| July | 744 | 0.20 | 1.002 |
| August | 744 | -0.40 | 0.996 |
| September | 720 | -7.30 | 0.927 |
| October | 744 | -1.80 | 0.982 |
| November | 720 | -4.40 | 0.956 |
| December | 744 | 1.90 | 1.019 |

Table 2. Optimization problem parameters.

| | | | | | |
|---|---|---|---|---|---|
| $P_{11}^{linemax}$ = 16 kW | $P_{23\ or\ 32}^{linemax}$ = 2 $kW$ | $P_1^{bat\,min}$ = -2 kW | $P_2^{bat\,max}$ = 1 kW | $S_3$ = 1 kW | $target_{3,m}$ = 650 $ |
| $P_{12\ or\ 21}^{linemax}$ =8 kW | $P_{33}^{line\max}$ = 2 kW | $P_2^{bat\,min}$ = -1 kW | $P_3^{bat\,max}$ = 0.5 kW | $\overline{EDR}_{z,m}$ = *500* $ | $target^{total}$ = 1050 $ |
| $P_{13\ or\ 31}^{linemax}$ = 4 kW | $SOC_z^{min}$ = 0.2 | $P_3^{bat\,min}$ = -0.5 kW | $S_1$ = 4 kW | $target_{1,m}$ = 150 $ | $fm$ = 12 |
| $P_{22}^{linemax}$ = 4 kW | $SOC_z^{max}$ = 1 | $P_1^{bat\,max}$ = 2 kW | $S_2$ = 2 kW | $target_{2,m}$ = 250 $ | $fz$ = 3 |

As seen in Table 2, the total *target* of the system for the profit in each month is 1050 $. In addition, SMGs buy and sell power among themselves and with the main grid considering the mentioned maximum line capacities. In order to optimize the charge and discharge schedule of the batteries, the state of charge's minimum and maximum limits, power minimum and maximum limits, and base power values listed in Table 2 are utilized. On-peak hours refer to the hours beginning at 7:00 a.m. until 11:00 p.m. on weekdays, and off-peak hours that are between 11:00 p.m. and 7:00 a.m. on weekdays and all day on Saturdays, Sundays. On-peak and off-peak hours are used in the stylized short-run demand model for electricity. The hourly price of the electricity in each month is illustrated in Figure 6. These illustrations are based on the formulations mentioned in *Section 3.1.* The same price is used for with and without COVID-19 cases to be able to solely assess the effect of COVID-19 on the system. This is a fair assumption because we use a range from -0.16 to -0.08 for elasticity. As discussed in *Section 3.1.*, considering the studies performed in the U.S. in [33], the average elasticity values in individual studies are below 1 which shows that the percentage change in the quantity demanded is less than the percentage change in price. This means that consumers are relatively unresponsive to price change.

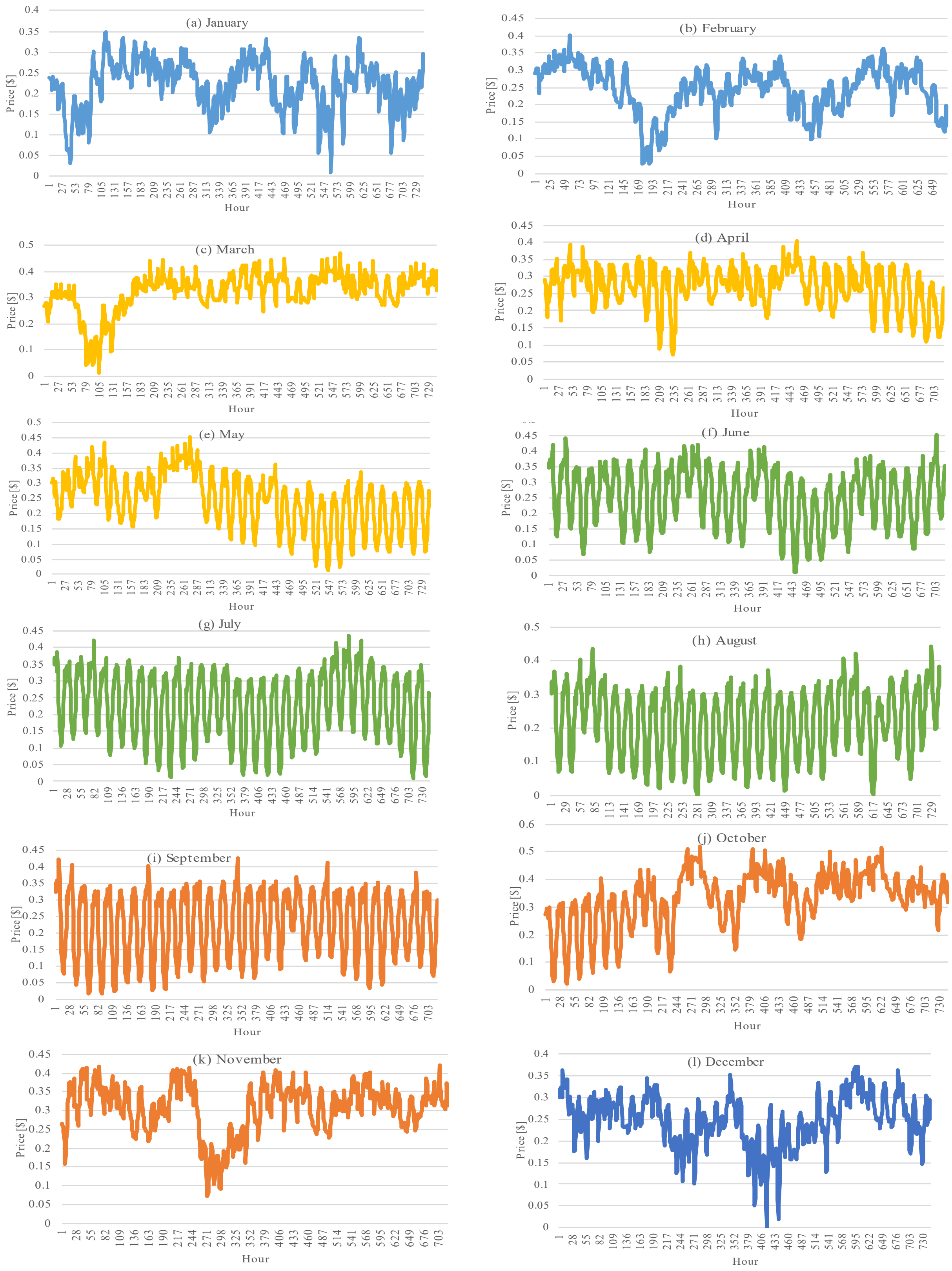

Figure 6. Hourly power price for 12 months.

As shown in Figure 6, the price of electricity varies due to the change in the demand of the consumers. The demand is affected by several factors including COVID-19, lockdown, seasons, weather, holidays, events, and so on. In this figure, each subfigure is shown with a specific color regarding its season to demonstrate the similarity of electricity prices within each season; spring, summer, fall, and winter are illustrated by yellow, green, orange, and blue, respectively. As seen in Figure 6, the subfigures with the same color look similar to each other considering the ambient conditions. It should be noted that when SMGs sell power to the main grid the price is 10% more than what is shown in Figure 6 due to the 10% bonus that the main grid pays to the SMGs. Also, the price of the sale or purchase of power inside the system is the same.

By performing the simulations of the first and second scenarios, the profit of the MMG system (see Figure *1*) with and without considering the effect of the COVID-19 pandemic is obtained. Figure 7 shows the profit of each of the SMGs and the collective profit of the MMG for 12 months. As seen, in January, March, April, May, August, September, October, and November the profit of the MMG and all SMGs are improved due to the COVID-19 and reduction of the demand. On the other hand, in February, June, July, and December the profit of the system is decreased due to the increase in the demand of the consumers. Also, the improvement of the profit in September is remarkable. Since the demand dropped by 7.3% because of the COVID-19 effect, the profit is increased significantly. Despite the fact that the demand is slightly decreased in August (0.4%), the profit of the system is still negative as the system could not supply its demand completely and it purchased power from the main grid.

Additionally, the results of the third and fourth scenarios are shown in Figure 8. As seen, the trend of the profit changes in the third and fourth scenarios are the same as the first and second scenarios; the profit improved in eight months and decreased in four months due to the effect of the demand. Like the first and second scenarios, September has the highest improvement among all twelve months. However, there are some important differences between Figure 7 and Figure 8 that should be taken into consideration. Compared to Figure 7, Figure 8 which represents the system in Figure 2, has two more months (July and September) in addition to August with negative profit for SMGs and MMG. This shows that the system is in shortage of power to supply its demand during these three months. In September, COVID-19 has made the profit of the system positive considering the high decline in the demand, which is not the case in July and August.

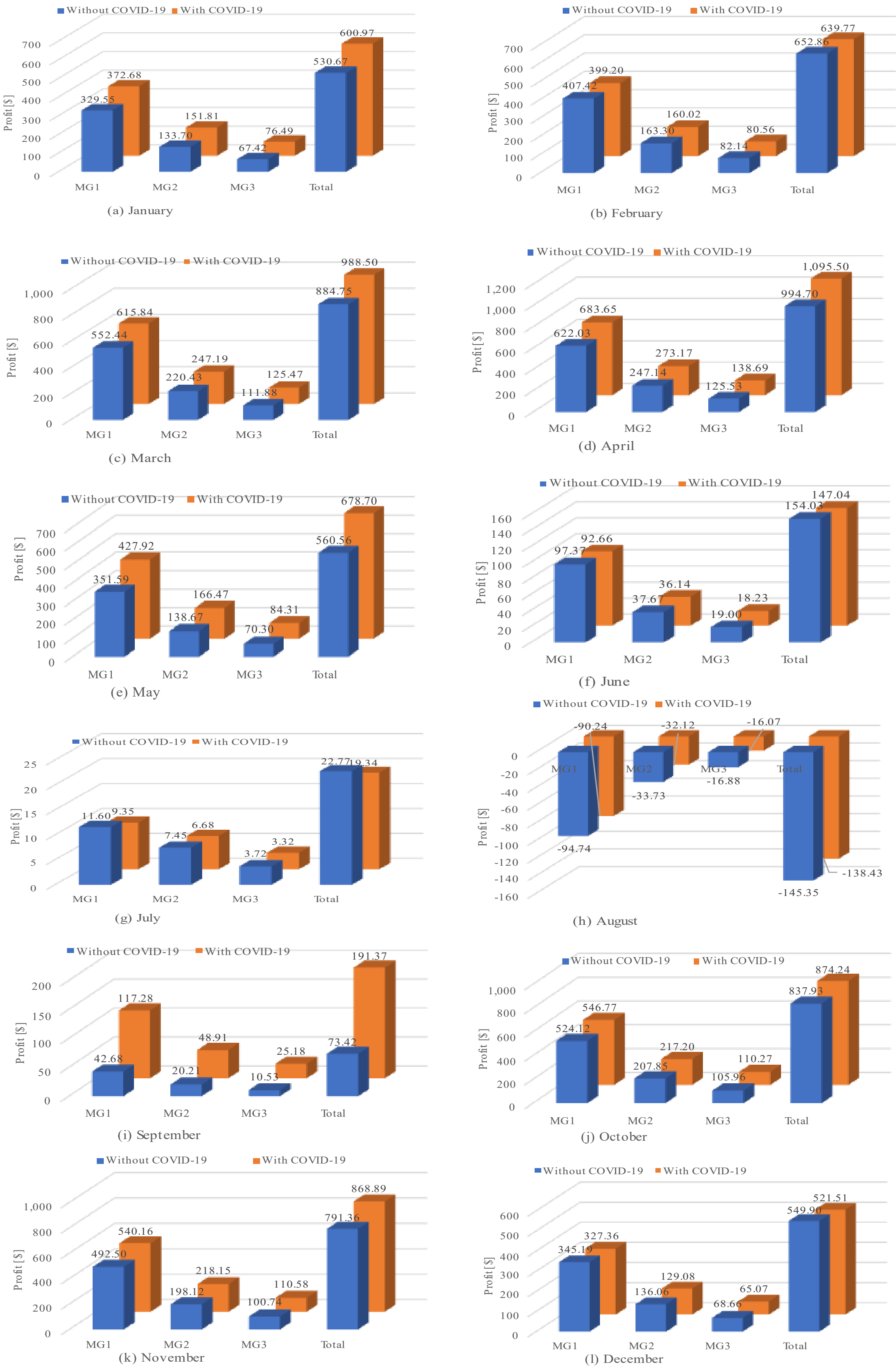

Figure 7. Profit of the MMG with and without COVID-19 for 12 months (first and second scenarios).

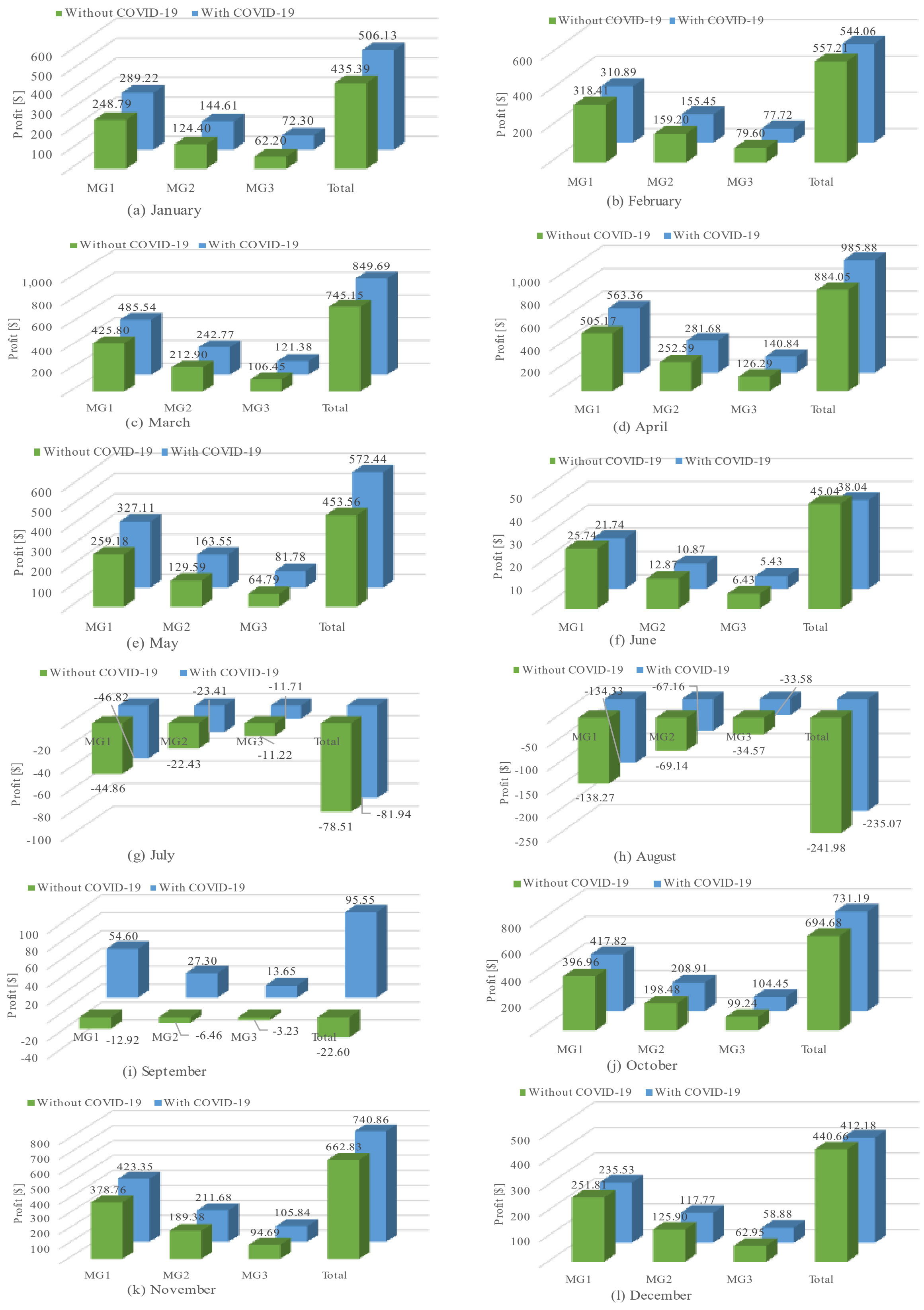

(a) January (b) February (c) March (d) April (e) May (f) June (g) July (h) August (i) September (j) October (k) November (l) December

Figure 8. Profit of the SMGs with and without COVID-19 for 12 months (third and fourth scenarios).

A radar chart of the average monthly price of power is shown in Figure 9. This figure highlights the price fluctuation over a year. As seen, the system has its least price in August and its highest price in October that is in line with the proposed demand model. Also, the outputs of the objective functions during a year are compared in Figure 10 considering all four scenarios. The system has the maximum profit in April and the minimum profit in August.

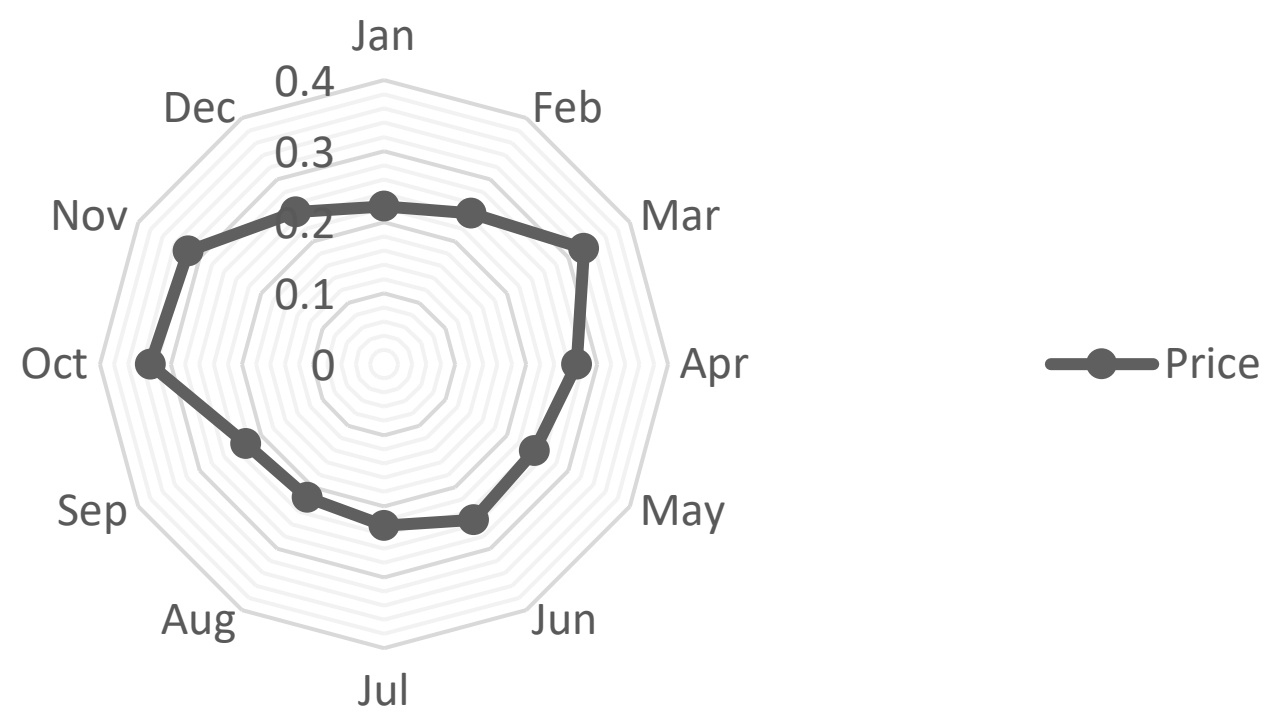

Figure 9. Average price of the power throughout the year [$].

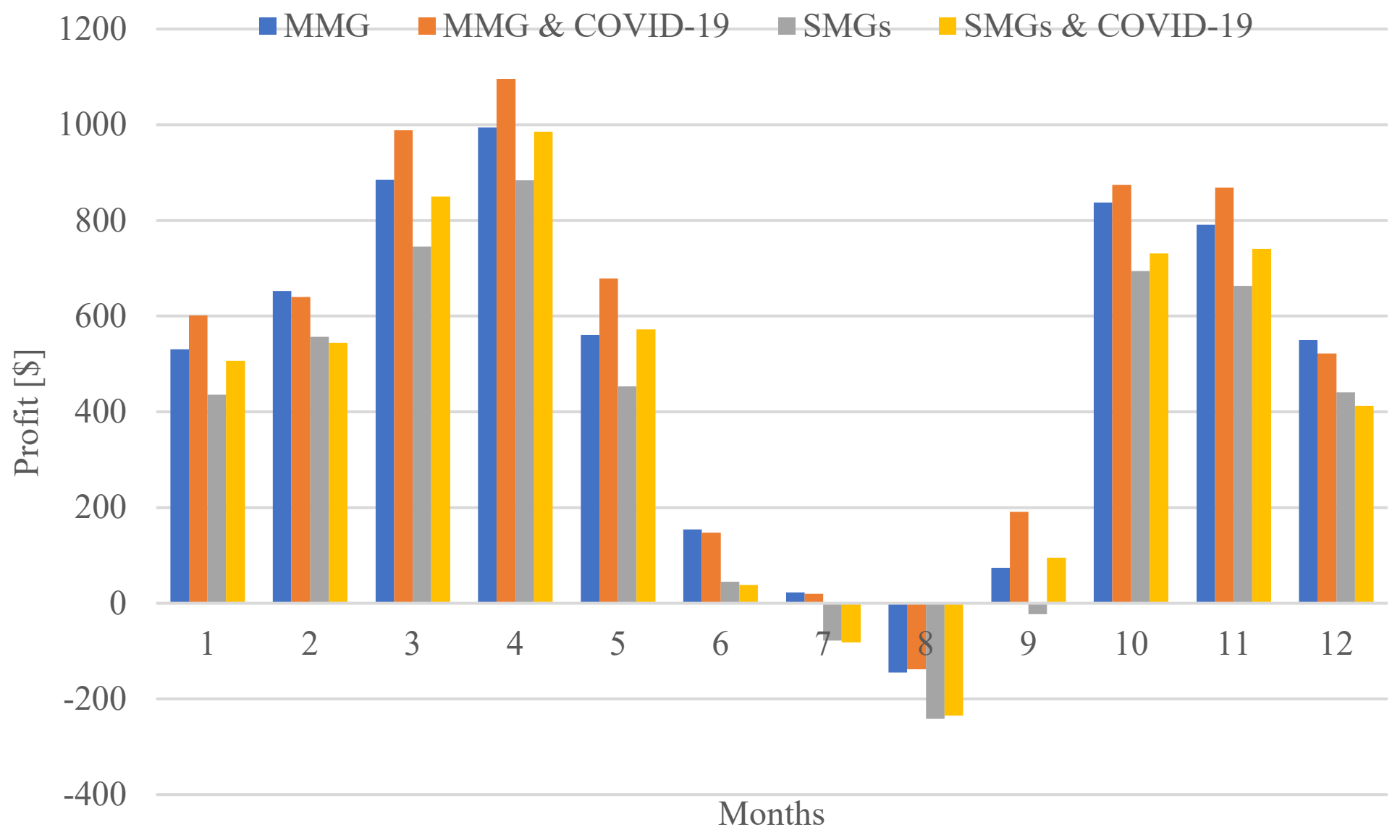

Figure 10. Comparison of the four scenarios.

In Figure 11, the yearly aggregate profit for all scenarios is illustrated. By comparing the first and third scenarios, it is observed that the profit of the system is increased by 25.57% by forming an MMG as well as the presence of the COVID-19's effect. Moreover, the comparison between the second and fourth scenarios indicates that clustering the SMGs enhances the profit of the system by 29.11%, which is drastically a high number. This is due to the trading power among different SMGs when one SMG needs power and the other has excess power.

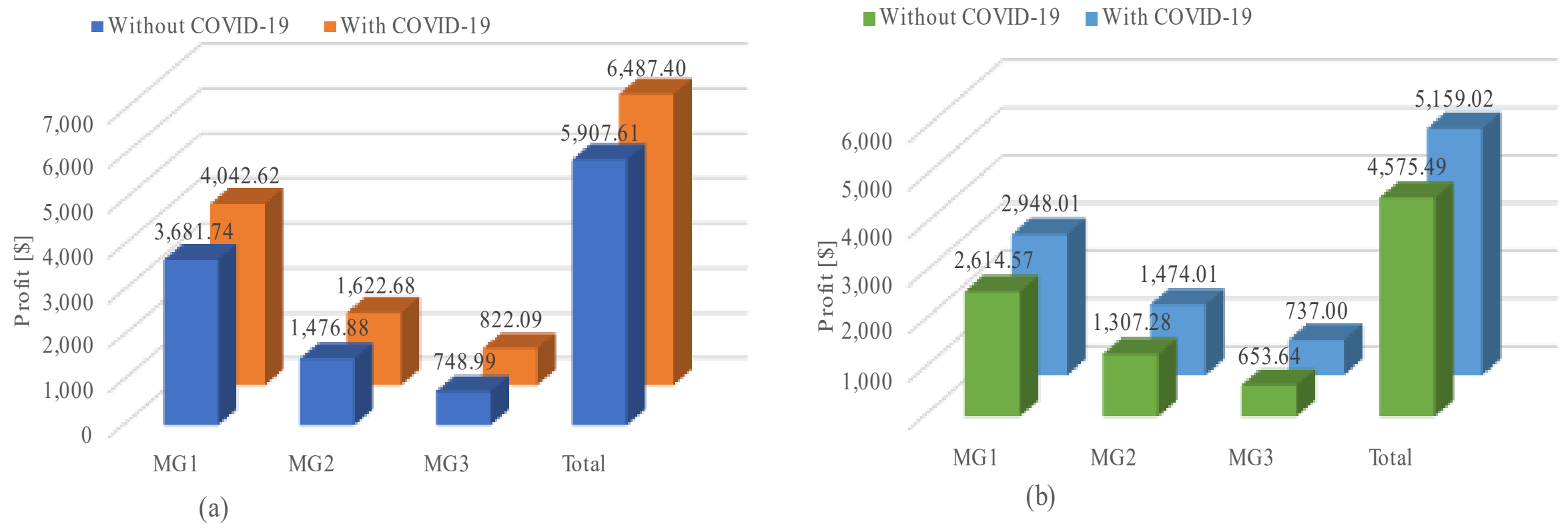

Figure 11. Annual total profit of MMG and SMGs comparison with and without COVID-19 (all scenarios).

Table 3 demonstrates the percentage of the total profit change due to the COVID-19's effect in the GC SMGs and GC MMG during a year. As seen, the COVID-19 increased the profit of the SMGs and MMG by about 12.75% and 9.81 %. In Table 4, the risk of the system in the first and second scenarios is analyzed. This table shows that the MMG had a decrease of 7.814% in its risk due to the COVID-19's impact. In Table 5, the risk of the GC SMGs in the third and fourth scenarios are tabulated. The indicated results show that the SMGs are experienced a 6.880% reduction in their risk regarding the effect of the COVID-19 pandemic over a year. Furthermore, the clustering of SMGs resulted in a 16.627% and 17.440% reduction in the system's risk with and without COVID-19, respectively. It is worth saying that the EIA expects that reduction of the COVID-19-related restrictions and economic development will cause more energy use in 2021 [35].

As a side note, the system has to include the energy storage systems to make sure to support the system during low irradiation and low wind speed considering the local nature of the system. The payback period for a system like the considered one that uses PVs located in New Mexico is estimated between 4.74- 25.68 years depending on the system size and the power price, etc. The payback period for WTs is estimated between 13-19 years depending on the size as well as other factors.

Table 3. Profit change due to COVID-19 effect [%].

| # | Jan | Feb | Mar | Apr | May | Jun | Jul | Aug | Sep | Oct | Nov | Dec | Year Average |
|---|---|---|---|---|---|---|---|---|---|---|---|---|---|
| **Total Profit SMGs** | 16.25 | -2.36 | 14.03 | 11.52 | 26.21 | -15.55 | -4.37 | 2.86 | 522.72 | 5.26 | 11.77 | -6.46 | 12.75 |
| **Total Profit MMG** | 13.25 | -2.01 | 11.73 | 10.13 | 21.07 | -4.54 | -15.06 | 4.76 | 160.63 | 4.33 | 9.80 | -5.16 | 9.81 |

Table 4. Risk change of MMG due to COVID-19 effect [%].

| # | Jan | Feb | Mar | Apr | May | Jun | Jul | Aug | Sept | Oct | Nov | Dec | Year Average |
|---|---|---|---|---|---|---|---|---|---|---|---|---|---|
| **MG1** | -13.460 | 3.388 | -64.982 | -100 | -25.578 | 0.851 | 0.352 | -0.605 | -12.282 | -17.993 | -30.263 | 5.849 | -7.946 |
| **MG2** | -15.568 | 3.790 | -90.495 | -100 | -24.972 | 0.720 | 0.321 | -0.568 | -12.490 | -22.180 | -38.600 | 6.131 | -8.052 |
| **MG3** | -10.983 | 2.336 | -35.653 | -53.780 | -17.575 | 0.584 | 0.277 | -0.484 | -10.505 | -9.787 | -19.972 | 4.407 | -6.955 |
| **Total** | -13.538 | 3.296 | -62.782 | -79.541 | -24.137 | 0.781 | 0.334 | -0.579 | -12.077 | -17.121 | -29.975 | 5.678 | -7.814 |

Table 5. Risk change of SMGs due to COVID-19 effect [%].

| # | Jan | Feb | Mar | Apr | May | Jun | Jul | Aug | Sept | Oct | Nov | Dec | Year Average |
|---|---|---|---|---|---|---|---|---|---|---|---|---|---|
| **MG1** | -10.076 | -1.323 | -26.644 | -40.179 | -17.381 | 0.641 | 0.282 | -0.501 | -10.185 | -8.244 | -16.439 | 4.087 | -6.427 |
| **MG2** | -16.093 | -1.797 | -80.509 | 0.000 | -28.208 | 0.844 | 0.360 | -0.619 | -13.163 | -20.245 | -36.778 | 6.557 | -8.003 |
| **MG3** | -11.510 | -1.451 | -34.292 | -61.366 | -19.932 | 0.697 | 0.304 | -0.535 | -11.016 | -10.274 | -20.155 | 4.674 | -7.227 |
| **Total** | -11.510 | -1.450 | -34.292 | -43.159 | -19.931 | 0.697 | 0.304 | -0.535 | -11.016 | -10.274 | -20.155 | 4.674 | -6.880 |

## Conclusion

COVID-19 pandemic has had a significant influence on people's lives in a variety of forms. COVID-19's breakout has resulted in economic hardship and massive waves of mortality. That said, the COVID-19 pandemic has an incredibly important effect on the power systems as well. In the U.S., the total consumed power has been declined by about 2.88%. during 2020 compared to 2019. This paper performs an annual comprehensive analysis of the COVID-19 pandemic’s effect on the profit and risk of the power systems by proposing three different SMGs. Different perspectives of the economic changes have been assessed considering the load profile change during several months. The demand of the three presented GC SMGs is mainly supplied by their own WTs, PVs, and batteries. In addition, the clustering of the SMG to form MMG during COVID-19 is also considered. The objective function of the research is profit maximization considering the trade between the main grid and different parts of the studied system. Moreover, the downside risk constraints are employed to find the risk of the system in different architectures. As a result, an MIP problem is proposed to find the mentioned variables. Furthermore, a stylized short-run heterogeneous consumers demand model is used regarding the on-peak and off-peak hours of the load profile. Since the COVID-19 reduces the demand for the power system, the profit of both SMGs and MMGs when considering the pandemic’s effect are increased by 12.75% and 9.81%, respectively. The obtained simulation results also show that the decrement of the risk at SMGs and MMGs are 6.88% and 7.814%, respectively. By clustering the SMGs and forming the MMG, the profit of the system has an increase of 29.11% and 25.75% with and without considering the COVID-19 pandemic, respectively. Moreover, the clustering of SMGs resulted in a 16.627% and 17.440% reduction in the system's risk with and without COVID-19, respectively. Therefore, it is noticeable that this pandemic can help the power systems to better prepare for future possible worldwide crises considering its impact on the power systems. In addition, it is beneficial to cluster the SMGs in order to create MMGs even during pandemics. By using several scenarios and architectures, authors have tried to validate the proposed model using the real input. Authors believe that this proposed system can be used and trusted regarding future happenings based on the obtained simulation results. As future work, evaluation of the proposed system's performance in the islanded mode can be investigated to exam the batteries' role in this model. Also, the system can be matured using other technologies. This work has considered the most well-known and popular renewable sources. Biomass technology will be used in a future work due to its environmental benefits. It should be mentioned that authors are reluctant to use diesel generators.

## Declaration of Competing Interest

The authors declare that they have no known competing financial interests or personal relationships that could have appeared to influence the work reported in this paper.

## Acknowledgment

This material is based upon work supported by the National Science Foundation EPSCoR Program under Award #OIA-1757207.